\documentclass[fleqn,10pt]{wlscirep}
\usepackage[utf8]{inputenc}
\usepackage[T1]{fontenc}
\usepackage{comment}

\title{Searching for Pneumothorax in X-Ray Images Using Autoencoded Deep Features}

\author[1]{Antonio Sze-To}
\author[1]{Abtin Riasatian}
\author[1,2]{H.R. Tizhoosh*}
\affil[1]{Kimia Lab, University of Waterloo, Waterloo, ON, Canada N2L 3G1}
\affil[2]{Vector Institute, MaRS Centre, Toronto, ON, Canada M5G 1M1}
\affil[*]{\emph{Corresponding author: tizhoosh@uwaterloo.ca}}

\begin{abstract}
Fast diagnosis and treatment of pneumothorax, a collapsed or dropped lung, is crucial to avoid fatalities. Pneomothorax is typically detected on a chest X-ray image through visual inspection by experienced radiologists.
However, the detection rate is quite low. As well, there may be a shortage of highly specialized radiologists. Therefore, there is a strong need for an automated detection systems to assist radiologists. Currently, many deep learning solutions are subject to research and development in medical imaging. Despite the high accuracy levels generally reported for deep learning classifiers in many applications, they may not be useful in clinical practice due to lack of large high-quality labelled image sets as well as lack of interpretation possibility. Alternatively, searching in the archive of past cases to find matching images may serve as a ``virtual second opinion'' through accessing the metadata of matched evidently diagnosed cases. To use image search as a triaging or diagnosis assistant, all chest X-ray images must first be tagged with expressive identifiers, i.e., deep features. Then, given a query chest X-ray image, the majority vote among the top $K$ retrieved images can provide a more explainable output.
While image search can be clinically more viable, its detection performance needs to be investigated at a scale closer to real-world practice.
In this study, we combined three public datasets to assemble a repository with more than 550,000 chest X-ray images. We developed the Autoencoding Thorax Net (short \emph{AutoThorax}-Net) for image search in chest radiographs compressing three inputs: the left chest side, the flipped right side, and the entire chest image. Experimental results show that image search based on \emph{AutoThorax}-Net features can achieve high identification performance providing a path towards real-world deployment. We achieved $92\%$ AUC accuracy for a semi-automated search in 194,608 images (pneumothorax and normal) and $82\%$ AUC accuracy for fully automated search in 551,383 images (normal, pneumothorax and many other chest diseases).
\end{abstract}
\begin{document}
\flushbottom
\maketitle

\thispagestyle{empty}

\section*{Introduction}
\begin{figure*}[htb]
\centering
\includegraphics[width=0.45\textwidth]{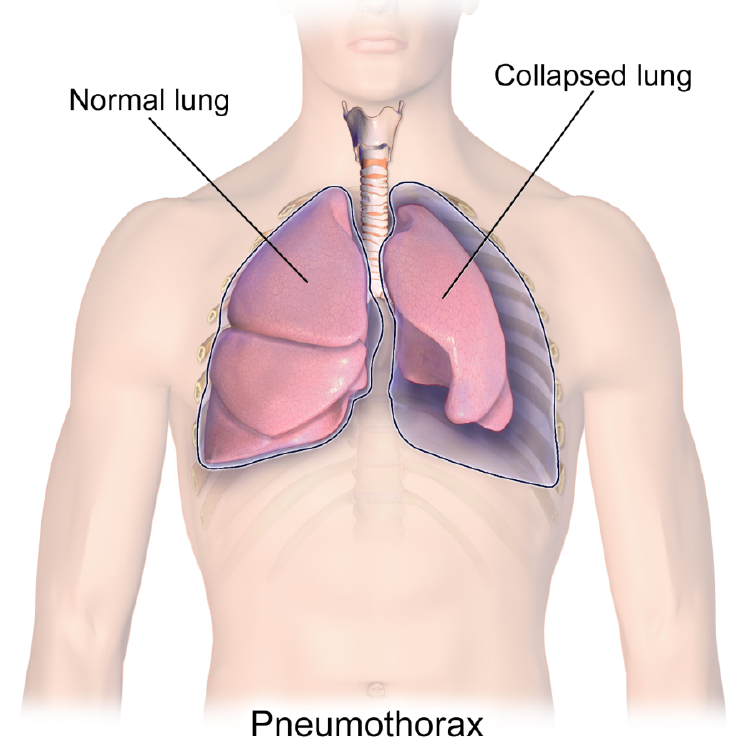}
\caption{A graphical illustration of pneumothorax. [Medical gallery of Blausen Medical 2014, WikiJournal of Medicine \cite{richfield2014medical}]}
\label{fig:demo}
\end{figure*}

Pneumothorax (collapsed or dropped lung) is an emergency condition when air enters the pleural space, i.e., the space between the lungs and the chest wall \cite{imran2017pneumothorax,zarogoulidis2014pneumothorax}.
A graphical illustration of pneumothorax is provided in Figure  \ref{fig:demo}.

It is generally a serious condition that can be fatal \cite{imran2017pneumothorax}.
To prevent patient death, early detection of pneumothorax through application of deep learning may be a viable option  \cite{goossen2019deep}. Pneumothorax is typically detected on chest X-ray images by qualified radiologists \cite{zarogoulidis2014pneumothorax}.
However, nowadays radiologists in many may have to process many X-ray studies daily \cite{guendel2018learning}.
With the increasing workload, the large volume of work for radiologists understandably delays diagnosis and treatment. In this process, experience is absolutely necessary but  
even the most experienced expert may be prone to miss the subtleties of an image  \cite{kelly2016development,mao2018deep}.
Since a wrong or delayed diagnosis can cause harm to patients \cite{ker2017deep}, it is vital to develop computer-aided approaches to assist radiologists in their daily workflow.


Due to its recent success, an increasing number of studies have adopted ``\emph{deep learning}'' for processing digital images, referring to the use of Deep Neural Networks (DNN), defined as artificial neuronal networks with more 3 hidden layers \cite{bengio2009learning} (DNNs practically consisting of many more hidden layers), to detect pneumothorax or other thoracic diseases in chest X-ray images \cite{wang2017chestx, rajpurkar2017chexnet, taylor2018automated, rubin2018large, rajpurkar2018deep, dunnmon2018assessment, feng2019deep, irvin2019chexpert, baltruschat2019comparison, rakshit2019deep, sze2019tchexnet, lecun2015deep}.
The deep-learning  pneumothorax detection systems can be categorized into two categories: 1) detection methods, i.e., to pinpoint the exact location of certain thoracic diseases in an input chest X-ray image, and 2) classification methods, i.e., to identify the presence of certain thoracic diseases in an input chest X-ray image without highlighting the exact location of the disease.

From a practical point of view, detection systems remain hard to be developed as large high-quality datasets with pixel-level labels are needed to train such systems. These datasets are expensive to obtain as creating representative and accurate labels constitutes tedious manual work for radiologists (for instance, many works focus on large and medium size pneumothorax \cite{taylor2018automated}). Classification systems, on the other hand, are relatively easier to develop as they only need image-label annotation. 

So far, more than half a million chest X-ray images with image-level (or global) labels have been released. These are ChestX-ray14 \cite{wang2017chestx}, CheXpert \cite{irvin2019chexpert} and MIMIC-CXR \cite{johnson2019mimic}.
Leveraging such a large amount of labelled images, classification-based systems should not be difficult to train and deploy.
Nevertheless, the main drawback of a classification system is that it only outputs a single probability, a number that quantifies the likelihood of the chest X-ray to contain a certain abnormality. This may not be enough to justify the diagnosis.  

\textcolor{black}{Image search, as a different approach to classification, not only can provide a probabilistic output like a classifier by taking a weighted majority vote among matched images but also can provide access to the metadata of similar cases from the past, a functionality that a trained deep network for classification cannot offer. Hence, image search allows a comparison of patient histories and treatment profiles due to providing a list of matched cases and not just delivering a classification probability  \cite{zhang2014towards}}. Therefore, image search may enable a virtual ``second opinion'' for diagnostic purposes and provide computerized explanations for the decision support.
While image search may be more viable for clinical deployment in terms of explainability, its classification performance still needs to be investigated, specifically, whether image search can achieve a identification performance as high as those obtained by classification-based deep learning systems.

In this study, we explored the use of image search based on deep features obtained from DNNs to detect pneumothorax in chest X-ray images. By means of using image search as a classifier, all chest X-ray images were first tagged with a feature vector. Given a query chest X-ray image, the majority voting of the top $K$ retrieved chest X-ray images was then used as a classifier. The corresponding reports and other clinical metadata of the top search results can also be used if available. This is an inherent benefit of image search.

Our contributions in this study are two-folded.
Firstly, we developed \emph{AutoThorax}-Net that generates feature vectors from integrating multiple images into one feature vector. \textcolor{black}{Although the benefits of using deep features for processing x-ray images is  well-established, in this study we demonstrated breaking down the image into multiple sub-images, here by using the chest symmetry, does in fact provide better results; the separation of left and right lung accompanying the entire image apparently increases the recognition accuracy. Flipping one lung  -- as a coarse type of image registration -- may also contribute to the better feature matching.} Experimental results demonstrate that image search  on \emph{AutoThorax}-Net features can achieve a higher detection performance compared with using feature vectors solely from the input image.
Experimental results also showed that image search based using \emph{AutoThorax}-Net features with the dimensionality reduced by a factor of 12 times can achieve a detection performance, comparable to or outperforming those obtained by existing systems such as CheXNet \cite{wang2017chestx}. 

\section*{Related Works}

\subsection*{Pneumothorax and Chest X-Ray Images}
Pneumothorax often constitutes a medical emergency since the presence of air within the pleural cavity outside the lung leading to respiratory distress, particulary in critically ill patients \cite{imran2017pneumothorax, zarogoulidis2014pneumothorax}. It is one of the diseases in the \emph{Category 1} findings that should be communicated to clinicians within minutes in order to take immediate actions to avoid fatalities as recommended by the American College of Radiology (ACR) \cite{larson2014actionable, goossen2019deep}.

Chest X-ray is the most common medical imaging modality with over 35 millions images taken every year in the U.S. alone \cite{kamel2017utilization}. 
X-ray images allow for inexpensive screening of several chest conditions including pneumothorax \cite{guendel2018learning}. 
Since hospital daily workloads may result in long queues for radiology images to be read, including images acquired overnight or images without any clinical pre-screening, an automated method of inspecting chest X-rays and prioritizing studies with potentially positive findings for rapid review may reduce the delay in diagnosing and treating pneumothorax \cite{taylor2018automated}.

\subsection*{Deep Learning for Analyzing Chest X-Ray Images}
Since the release of the \emph{ChestX-ray14} dataset \cite{wang2017chestx} by National Institute of Health, providing 112,120 frontal-view X-ray images of 30,805 unique patients labelled for 14 diseases (in which each image may  have multi-labels), an increasing number of studies have adopted DNNs to develop automated systems to detect diverse diseases on chest X-ray images \cite{rajpurkar2017chexnet, taylor2018automated, rubin2018large, rajpurkar2018deep, dunnmon2018assessment, feng2019deep, irvin2019chexpert, baltruschat2019comparison, rakshit2019deep, sze2019tchexnet}.
\emph{CheXNet}\cite{rajpurkar2017chexnet}, a DNN with DenseNet121 architecture \cite{huang2017densely}, has been trained on ChestX-ray14 dataset and achieved radiologist-level detection of pneumonia. Since then, many DNN architectures have been proposed for a variety of tasks ranging from localization \cite{rakshit2019deep}, lateral and frontal dual chest X-ray reading \cite{rubin2018large}, integration of non-image data in classification \cite{baltruschat2019comparison}, attention-guided approaches \cite{guan2018diagnose}, location-aware schemes \cite{guendel2018learning}, weakly-supervised methods  \cite{yao2018weakly,yan2018weakly} as well as generative models \cite{mao2018deep}.

For detecting pneumothorax, a recent study collected 13,292 frontal chest X-rays (3,107 with pneumothorax) to train a DNN to verify the presence of a large or moderate-sized pneumothorax. Another recent study \cite{goossen2019deep} collected 1,003 images (437 with pneumothorax and 566 with no abnormality) to detect pneumothorax with DNNs. So far, there has been no study to investigate pneumothorax detection in a large dataset, perhaps by combining the three large public datasets.

\subsection*{Deep Learning for Content-Based Image Retrieval}
Retrieving similar images given a query image is known as Content-Based Image Retrieval (CBIR) \cite{zhou2017recent} or Content-Based Medical Image Retrieval (CBMIR) \cite{das2017overview} for medical applications. While classification-based methods provide promising results \cite{camlica2015medical}, CBMIR systems can assist clinicians by enabling them to compare the case they are examining with previous (already diagnosed) cases and by exploiting the information in corresponding medical reports \cite{das2017overview}. It may also help radiologists in faster and more reliably preparing reports for particular diagnosis \cite{das2017overview}.

While deep learning methods have been applied to CBIR tasks in recent studies \cite{wan2014deep, tzelepi2018deep}, there has been less attention on exploring deep learning methods for CBMIR tasks \cite{sklan2015toward, qayyum2017medical}.


One study investigated the performance of DNNs for MR and CT images with human anatomy labels \cite{sklan2015toward}. Another study investigated the retrieval performance of DNNs among multimodal medical images for different body organs \cite{qayyum2017medical}.
There is also a study exploring hashing deep features into binary codes, testing among lung, pancreas, neuro and urothelial bladder images \cite{qiu2017medical}.
Deep Siamese Convolutional Neural Networks \cite{chung2017learning} have also been tested for CBMIR to minimize the use of expert labels using multiclass
diabetic retinopathy fundus images. Another study explored deep learning for CBMIR among multiple modalities for a large number of classes \cite{owais2019effective}.
So far, there has not been any report to validate CBMIR techniques for a challenging case like pneumothorax detection in large datasets. We attempt to close this gap by reporting our results on a large dataset of chest x-ray images by fusion three public datasets. 

\section*{Methods}
Given a query chest X-ray image, the problem is to output whether it contains pneumothorax using image search in archived images through majority vote among retrieved similar images from the archive.

\begin{figure*}[htb]
  \centering
  \includegraphics[width=0.9\textwidth]{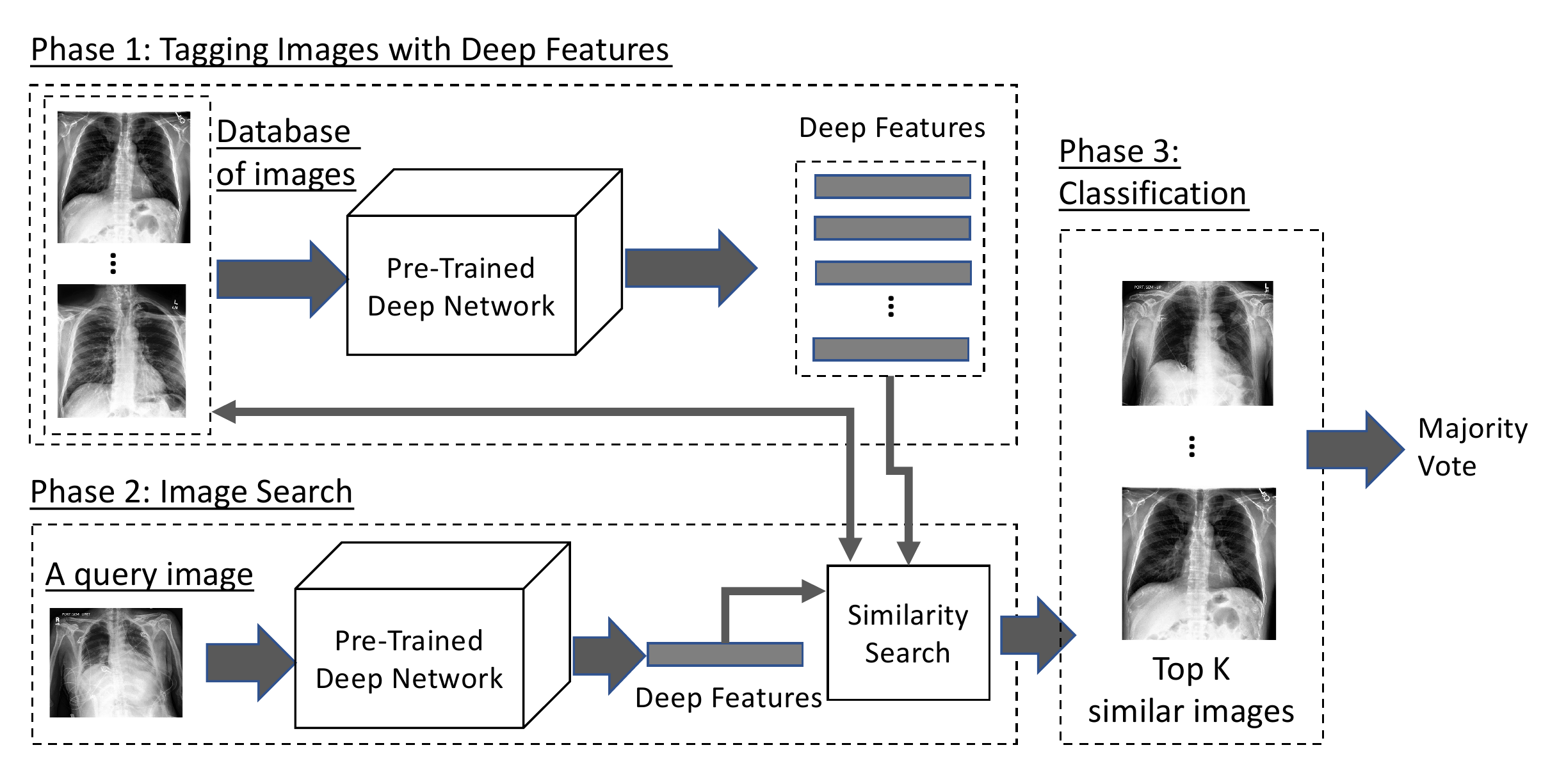}
  \caption{An overview of using image search as a classifier to recognize pneumothorax in chest X-ray images.  Phase 1: Tagging images with Features. Phase 2: Image Search (distance claculation between the query features and all other features in the database). Phase 3: Classification (the majority voting of the retrieved images as a classifier).
}
\label{fig:method}
\end{figure*}

The proposed method of using image search as a classifier comprises of three phases (Figure  \ref{fig:method}):
1) Tagging images with deep features (all images in the database are fed into a pre-trained network to extract features), 
2) image searching (tagging the query image with features and calculating its distance with all other features in the archive to find the most similar images), 
3) classification (majority vote among the labels of retrieved similar images).

\begin{figure*}[htb]
  \centering
  \includegraphics[width=0.8\textwidth]{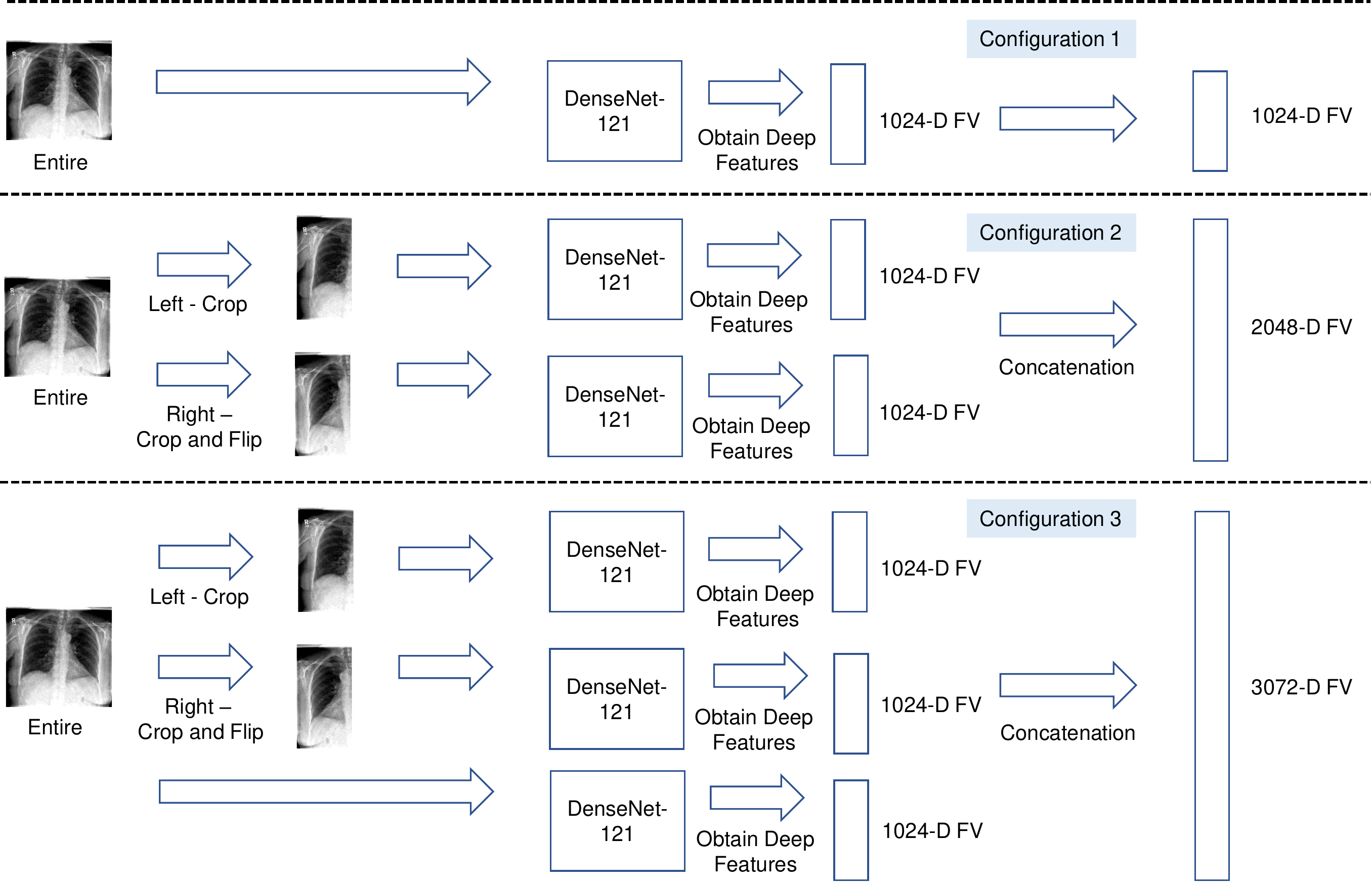}
  \caption{An overview of the three configurations using DenseNet121\cite{huang2017densely} to extract features from a chest X-ray images. 
Configuration 1: a feature vector is extracted from the entire chest X-ray image. Configuration 2: two feature vectors are extract from the left chest side and flipped right chest side. The final feature vector is a concatenation of these two feature vectors. Configuration 3: three feature vectors are extract from the left chest side, the flipped right chest side, and the entire chest X-ray image, respectively.
}
\label{fig:configuration}
\end{figure*}

\textbf{Phase 1: Tagging Images with Deep Features} -- In this phase, all chest X-ray images in the archive are tagged with deep features. To represent a chest X-ray image as a feature vector with a fixed dimension, the last pooling layer may be used as image represnetation. In other words, the pre-trained deep convolutional neuronal network is considered as a feature extractor to convert a chest X-ray image into an $n$-dimensional feature vector with $n=1,024$ being a typical value for many networks.
In our study, DenseNet121 \cite{huang2017densely} is used for converting a chest X-ray image into a feature vector with 1,024 dimensions. \textcolor{black}{DenseNet topology has a strong gradient flow contributing to diverse features and is, compared to many other architectures such as ResNet and EfficientNet, quite compact with \emph{only} 7 million weights. We adopted DenseNet121 also for a fair comparison in experiments with CheXNet, which has also used DenseNet121 as the architecture. } Three configurations are explored to extract deep features from a chest X-ray image (Figure \ref{fig:configuration}):

\begin{itemize}
    \item  \textbf{Configuration 1 --} a feature vector is extracted from the entire chest X-ray image. Representing the entire image with one feature vector is quite common and assumes that the object or abnormality will be adequately quantified in the single feature vector for the entire image. 

    \item \textbf{Configuration 2 --}  two feature vectors are extracted, one from the left chest side and one from the flipped version of the right chest side. The final feature vector is a concatenation of these two feature vectors. If DenseNet121 \cite{huang2017densely} is adopted as the feature extractor, the feature vector has $2,048$ values. The rational behind this idea is to allow expressive features for each side of the chest to be quantified separately to make feature matching easier for the unsupervised search. As well, flipping the right lung is a registration-like operation to facilitate alignment in matching.    

    \item \textbf{Configuration 3 --} three feature vectors are extracted as decsribed in previous two configurations.  The final feature vector is a concatenation of these three feature vectors. If DenseNet121 \cite{huang2017densely} is adopted as the feature extractor, the dimension of the final feature vector is $3,072$ real-valued features. The rationale behind this configuration is that matching a combined feature vector that represents the whole image, the left chest side and the flipped right chest side not only provides a  global image view but also more focused and aligned attention to each chest side to emphasize their features in the search and matching process. 

\end{itemize}

\textbf{Phase 2: Image Search} -- In this phase, the distance between the deep features of the query chest X-ray image and all chest X-ray images in the database are computed. The chest X-ray images having the shortest distance with those of the query chest X-ray image are subsequently retrieved. \textcolor{black}{The Euclidean distance, as the most commonly used norm for deep feature matching, was used for computing the distance between the deep features of two given chest X-ray images. It is the geometric distance in the multidimensional space recommended when all variables have the same metric \cite{mercioni2019survey}.} The calculated distances can be sorted to retrieve as many as matched images as desired. \textcolor{black}{The impact of distance norms on retrieval may be investigated in future works.} 

\textbf{Phase 3: Classification} -- 
In this phase, the majority vote among the labels of retrieved chest X-ray images is used as a classification decision. For example, given a query chest X-ray image, the top $K$ most similar chest X-ray images are retrieved. If $m$ chest X-ray images are labelled with pneumothorax (with $m\leq K$), the query image is classified as pneumothorax with a class likelihood of $m /k$. The larger $K$ the more reliable the classification vote will become. This, in turn, requires a large archive of tagged images to increase the probability of finding similar images.

\subsection*{Compressing Feature Dimensionality using Autoencoders}
\begin{figure*}[htb]
  \centering
  \includegraphics[width=0.7\textwidth]{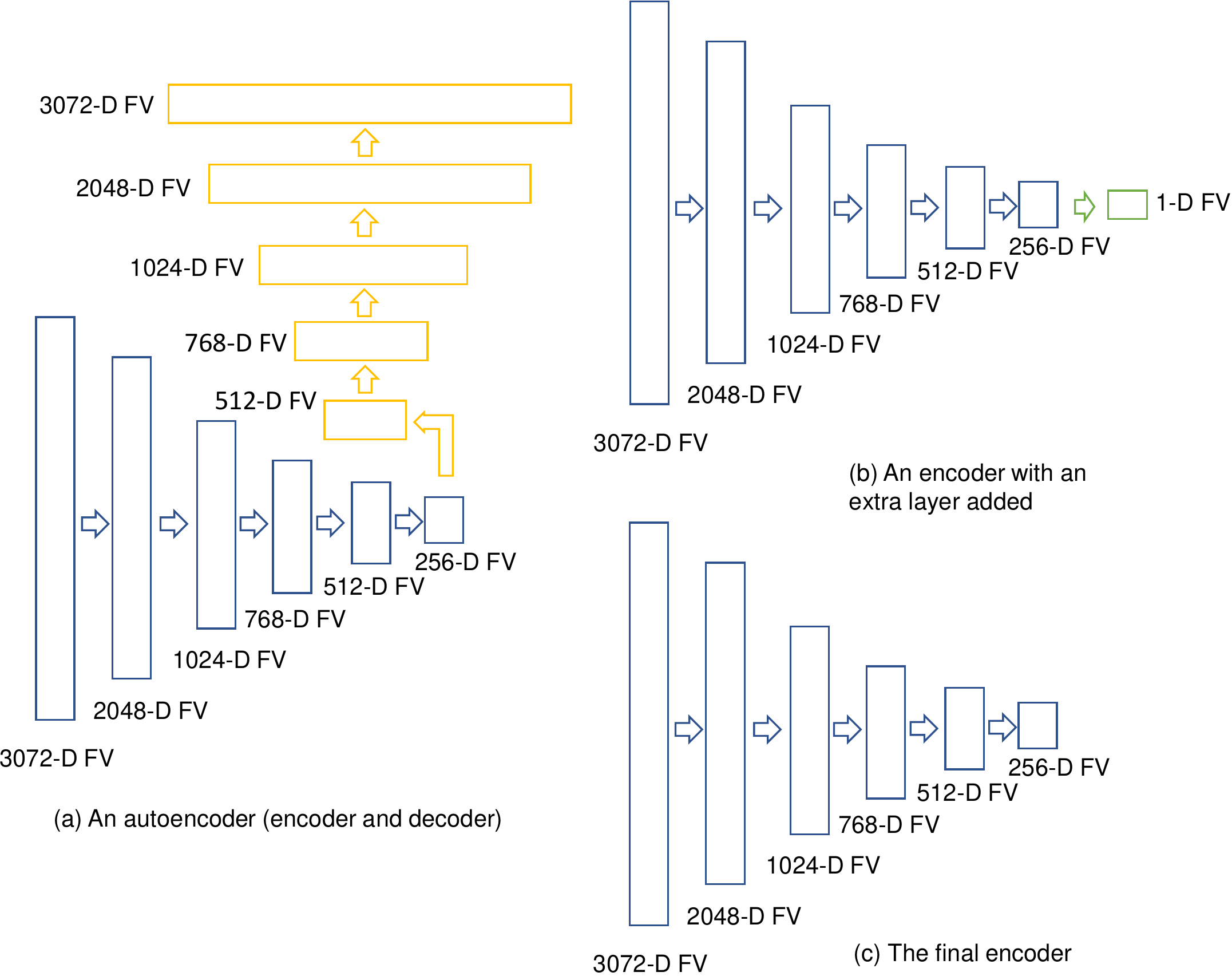}
  \caption{An overview of autoencoder topologies: 
(a) an autoencoder, with an encoder (highlighted in blue) and an decoder (highlighted in yellow), is constructed during step 1: Unsupervised end-to-end training with decoder,
(b) an encoder (highlighted in blue) with an extra layer added (highlighted in green) is constructed during step 2: supervised fine-tuning with labels, 
(c) an encoder (highlighted in blue) is constructed by removing the 1-dimension layer.
}
\label{fig:autoencoder}
\end{figure*}

The dimensionality of the feature vectors, especially the concatenated ones, may become a computational obstacle but it can be reduced by employing autoencoders. One may use an autoenoder for all configurations but our main motivation was a size reduction for the longest feature vector for configuration 3. Two steps are required to construct an encoder to reduce feature vector dimensionality:
\begin{itemize}
    \item \textbf{Step 1: Unsupervised end-to-end training with a decoder:} An autoencoder with the architecture summarized in Figure \ref{fig:autoencoder}(a) is first constructed. A dropout layer \cite{srivastava2014dropout} is introduced between each layer to reduce the probability of overfitting. The model is then trained by backpropagation with outputs being set equal to inputs.

    \item \textbf{Step 2: Supervised fine-tuning with labels:} After the training, the decoder in the model is removed as we only need the encoding part as dimensionality reduction. Instead, a one-dimensional fully connected layer of neurons with the sigmoid function as activation function was used in training phase.
\end{itemize}
The model architecture is summarized in Figure \ref{fig:autoencoder}(b).
Similarly, a dropout layer \cite{srivastava2014dropout} was introduced between the 256-dimensional layer and the one-dimensional layer to reduce the probabilty of overfitting. The model is then trained with through backpropagation. After the training, the one-dimensional fully connected layer is  removed. The model architecture of the final encoder is summarized in Figure \ref{fig:autoencoder}(c).

\subsection*{Model Architecture}
\begin{figure*}[htb]
\centering
\includegraphics[width=\textwidth]{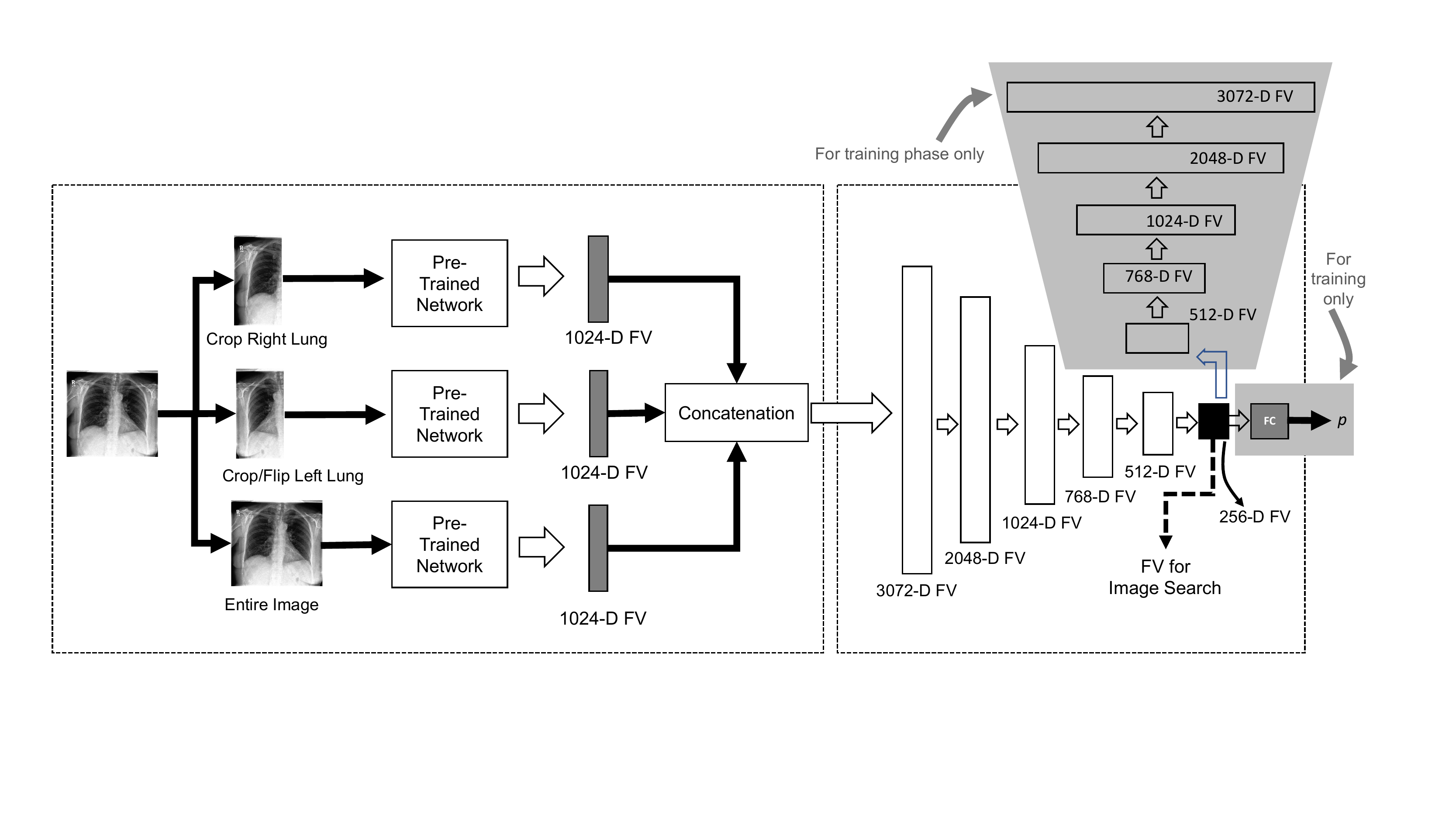}
\caption{A graphical illustration of \emph{AutoThorax}-Net. Three extracted feature vectors (FVs) (left) are concatenated and fed into into an encoder to compress them into 256 values for image search. During training the same compressed FV is used as input to a fully connected layer (FC) to classify images as pneumothorax with a likelihood of $p$.}
\label{fig:architecture}
\end{figure*}

The architecture of \emph{AutoThorax}-Net to obtain features from a chest X-ray image is illustrated in Figure  \ref{fig:architecture}.

\section*{Results}
In this section, we first describe the datasets collected and the preprocesing procedures.
We then describe the experiments that were conducted, followed by the analysis. \textcolor{black}{The main goal of experiments is to validate the performance of image search via matching deep features. In order to establish performance quantification, we treat search like a classifier by taking a consensus vote guided by the ROC statistics. We also compare the results with the CheXNet which is an end-to-end deep network specially trained for classifying chest X-ray images.}

\subsection*{Dataset Collection}
Three large public datasets of chest X-ray images were collected. 
The first is MIMIC-CXR \cite{goldberger2000physiobank, johnson2019mimic}, a dataset of 371,920 chest X-rays associated with 227,943 imaging studies. 
A total of 248,236 frontal chest X-ray images in the training set were used in this study.
The second dataset is CheXpert \cite{irvin2019chexpert}, a dataset  consisting of 224,316 chest radiographs belonging to 65,240 patients. A total of 191,027 frontal chest X-ray images in the training set were used in this study.
The third dataset is ChestX-ray14 \cite{wang2017chestx}consisting of 112,120 frontal-view X-ray images of 30,805 patients. All chest X-ray images in this dataset were used in this study.

In total, 551,383 frontal chest X-ray images were used in our investigations. 34,605 images (6\% of of all images) were labelled as pneumothorax. The labels refer to the entire image; the collapsed lungs were not highlighted in any way.

\subsection*{Implementation and Parameter Setting}
We used the library \emph{Keras} (\url{http://keras.io/}) v2.2.4 with Tensorflow backend \cite{abadi2016tensorflow} to implement the approach. As we used a pre-trained network for feature extraction, the DenseNet121 was selected \cite{huang2017densely}, and the weight file was obtained through the default setting of Keras. For CheXNet \cite{rajpurkar2017chexnet}, the  weight file was downloaded from GitHub (https://github.com/brucechou1983/CheXNet-Keras). 
All images were resized to 224$\times$224 before feeding into networks.
All other parameters were default values unless otherwise specified.
All experiments were run on a computer with 64.0 GB DDR4 RAM, an Intel Core i9-7900X @3.30GHz CPU (10 Cores) and one GTX 1080 graphic card.

\subsection*{Performance Evaluation}
Following relevant literature \cite{rajpurkar2017chexnet, irvin2019chexpert}, the performance of classification was evaluated by the area under the curve (AUC) for the receiver operating characteristic curve (ROC curve) to enable the comparison over a range of prediction thresholds.
As a 10-fold cross-validation was conducted in the experiments, average ROC was computed with 95\% confidence interval.

\subsection*{Dataset Preparation \& Preprocessing}
There is a concern for ChestX-ray14 \cite{wang2017chestx} dataset that its chest X-ray images with chest tubes were frequently labelled with Pneumothorax  \cite{zhang2019mitigating, zech2018variable}. As we combined ChestX-ray14 with CheXpert \cite{irvin2019chexpert}, and MIMIC-CXR \cite{goldberger2000physiobank, johnson2019mimic} datasets in our experiments, the concern was mitigated to address the bias.

\textbf{Dataset 1 (Semi-Automated Detection) --} This is a dataset comprising of 34,605 pneumothorax chest X-ray images and 160,003 normal chest X-ray images. Searching in this dataset means there is already a suspicion by the expert that the image may contain pneumothorax, hence the search is guided to only search within archived images that are diagnosed as either pneumothorax or normal (no finding).  
The pneumothorax images were obtained from the collected frontal chest x-ray images with the label "Pneumothorax".
They were considered as the positive (+ve) class.
The normal images were obtained from the collected frontal chest x-ray images with label the "No Finidng".
These chest X-ray images were considered as the negative (-ve) class.
A summary of dataset 1 is provided in Table \ref{tab:dataset1}.

\begin{table*}[htb]
    \centering
    \caption {A summary of chest X-ray images in the Dataset 1 through combination of three public datasets.} 
    \label{tab:dataset1} 
    \begin{tabular}{lcccc}
        \hline
        ~         & MIMIC-CXR \cite{johnson2019mimic,goldberger2000physiobank} & CheXpert \cite{irvin2019chexpert} & ChestX-ray14 \cite{wang2017chestx} & Total\\ \hline
        +ve: Pneumothorax & 11,610     & 17,693 & 5,302 & 34,605 \\   \hline
        -ve: Normal & 82,668     & 16,974 & 60,361 & 160,003  \\   \hline 
        Total & 94,278     & 34,667 & 65,663 & 194,608  \\   \hline
    \end{tabular}
\end{table*}

\textbf{Dataset 2 (Fully-Automated Detection) --} This dataset is comprising of 34,605 pneumothorax chest X-ray images and 516,778 non-pneumothorax chest x-ray images. Searching in this dataset means the computer is automatically searching in all images to verify the likelihood of pneumothorax without any guidance of the expert. 
The pneumothorax images were obtained from the collected frontal chest X-ray images with the label "Pneumothorax".
They were considered as the positive (+ve) class.
The non-pneumothorax images were obtained from the collected frontal chest X-ray images without the label "Pneumothorax", meaning that they may contain cases such as normal, pneumonia, edema, cardiomegaly and more. They were considered as the  negative (-ve) class.
A summary of dataset 2 is provided in Table \ref{tab:dataset2}.

\begin{table*}[htb]
    \centering
    \caption {A summary of chest X-ray images in the Dataset 2 through combination of three public datasets.} 
    \label{tab:dataset2} 
    \begin{tabular}{lcccc}
        \hline
        ~         & MIMIC-CXR \cite{johnson2019mimic,goldberger2000physiobank} & CheXpert \cite{irvin2019chexpert} & ChestX-ray14 \cite{wang2017chestx} & Total\\ \hline
        +ve: Pneumothorax & 11,610     & 17,693 & 5,302 & 34,605 \\   \hline
        -ve: Non-Pneumothorax & 236,626     & 173,334 & 106,818 & 516,778  \\   \hline 
        Total & 248,236     & 191,027 & 112,120 & 551,383  \\   \hline
    \end{tabular}
\end{table*}
\subsection*{First Experiment Series: Semi-Automated Solution}
\textcolor{black}{The first experiments series focuses on a ``semi-automated'' solution for pneumotharx. We confine the search and classification to cases that are either normal or diagnosed with pneumotharx (Dataset 1). We test all three configurations (Figure \ref{fig:configuration}), CheXNet, and the proposed \emph{AutoThorax}-Net.}

\subsubsection*{Experimental Workflow}
\begin{figure*}[htb]
\centering
\includegraphics[height=5.5cm]{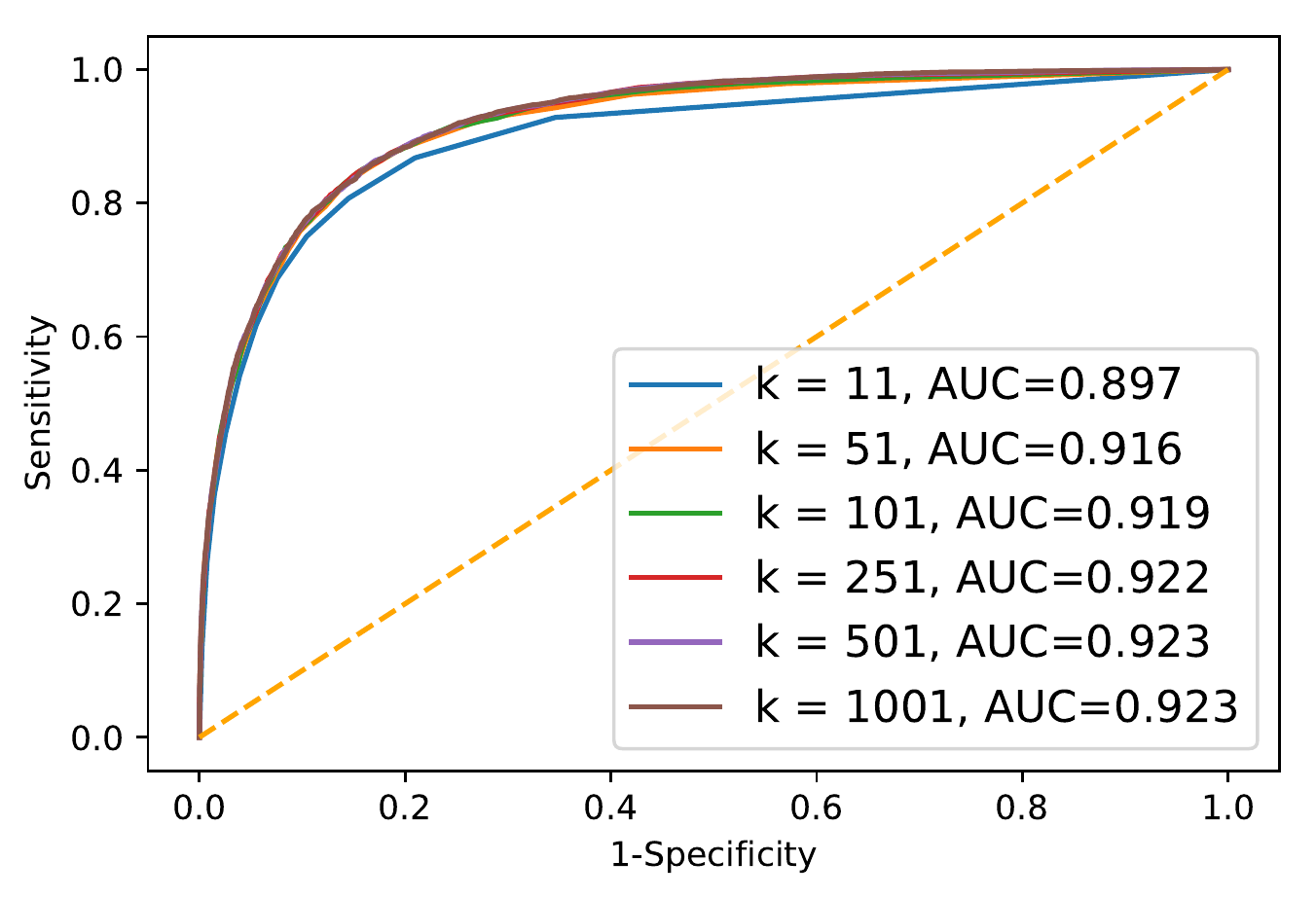}
\includegraphics[height=5.5cm]{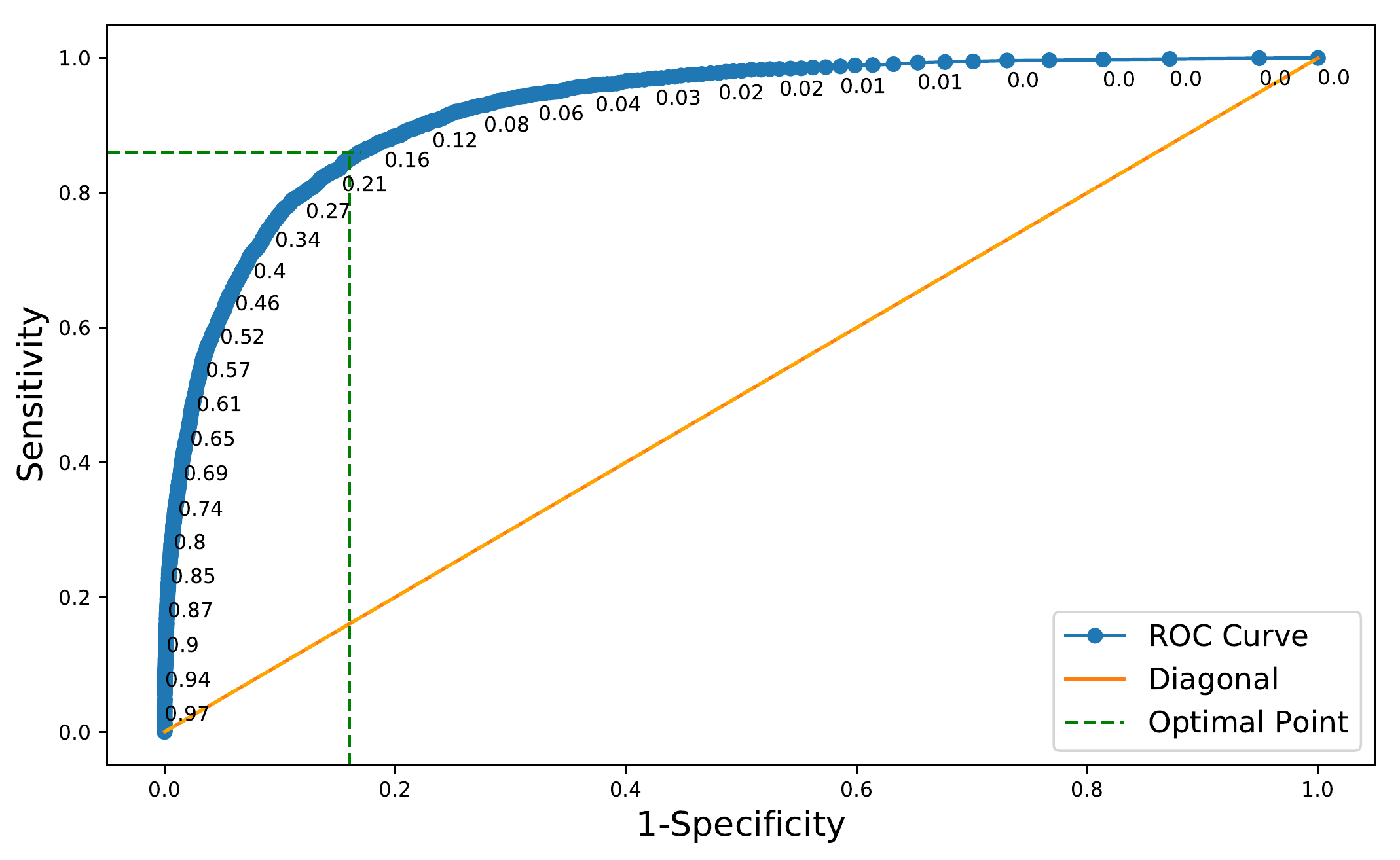}
\caption{\textcolor{black}{Analysis for Dataset 1. Left: Sample ROC curves for the proposed \emph{AutoThorax}-Net for different $k$ values in one fold and their area under the curve (AUC), Right:  corresponding ROC thresholds for $k=1001$ to select the sensitivity-specificity trade-off using Youden's index.}}
\label{fig:roc1}
\end{figure*}

All images of Dataset 1 for all configurations were first tagged with deep features. \textcolor{black}{We constructed the receiver operating characteristics (ROC) curve for the dataset to find the trade-off between sensitivity and specificity (Figure \ref{fig:roc1}). We used Youden's index \cite{youden1950index} to find the trade-off position on the ROC curve providing the threshold for match selection. The Youden's index can be calculated as  ``\emph{sensitivity}$+$\emph{specificity}$-1$''.}
A standard 10-fold cross-validation was adopted for tests \textcolor{black}{ that showed a very low standard deviation for all experiments apparently due to the large size of the datasets}. 
All tagged chest X-ray images were divided into 10 groups. 
In each fold, one group of chest X-ray images was used as validation set, while the remaining chest X-ray images were used as ``archived'' images to be searched.
The above process was repeated 10 times, such that in each fold a different group of chest X-ray images was used as the validation set. \textcolor{black}{In each fold, an encoder was trained using the archived set of that fold. The encoder was then used for compressing deep features for each chest X-ray image in the validation set.}

\textcolor{black}{The parameters of the encoder construction process are described as follows:\\
\textbf{Step 1 --} Unsupervised end-to-end training with decoder: The training epoch and batch size were set as 10 and 128, respectively.
The loss function was chosen as mean-squared-error. Adam optimizer \cite{kingma2014adam} was used. The dropout rate was set to 0.2, i.e., a probability of 20\% setting the neuron output as zero to counteract possible overfitting.\\
\textbf{Step 2 --} Supervised fine-tuning with labels: The loss function was chosen as binary cross-entropy. Other parameters remained the same as in Step 1.}

Given a query chest X-ray image from validation sets, image search was conducted on the archived set to retrieve $K$ similar images for each query image. The consensus vote among the top $K$ retrieved chest X-ray images subsequently determines whether the query image is pneumothorax. \textcolor{black}{Results were generated with $K\in\{11, 51, 101, 251, 501 , 1001\}$.} As we were using a large number of archived images, one was excepting to see better results for higher K values.

\subsubsection*{Results}
\begin{figure*}[hb]
\centering
\includegraphics[height=3.5cm]{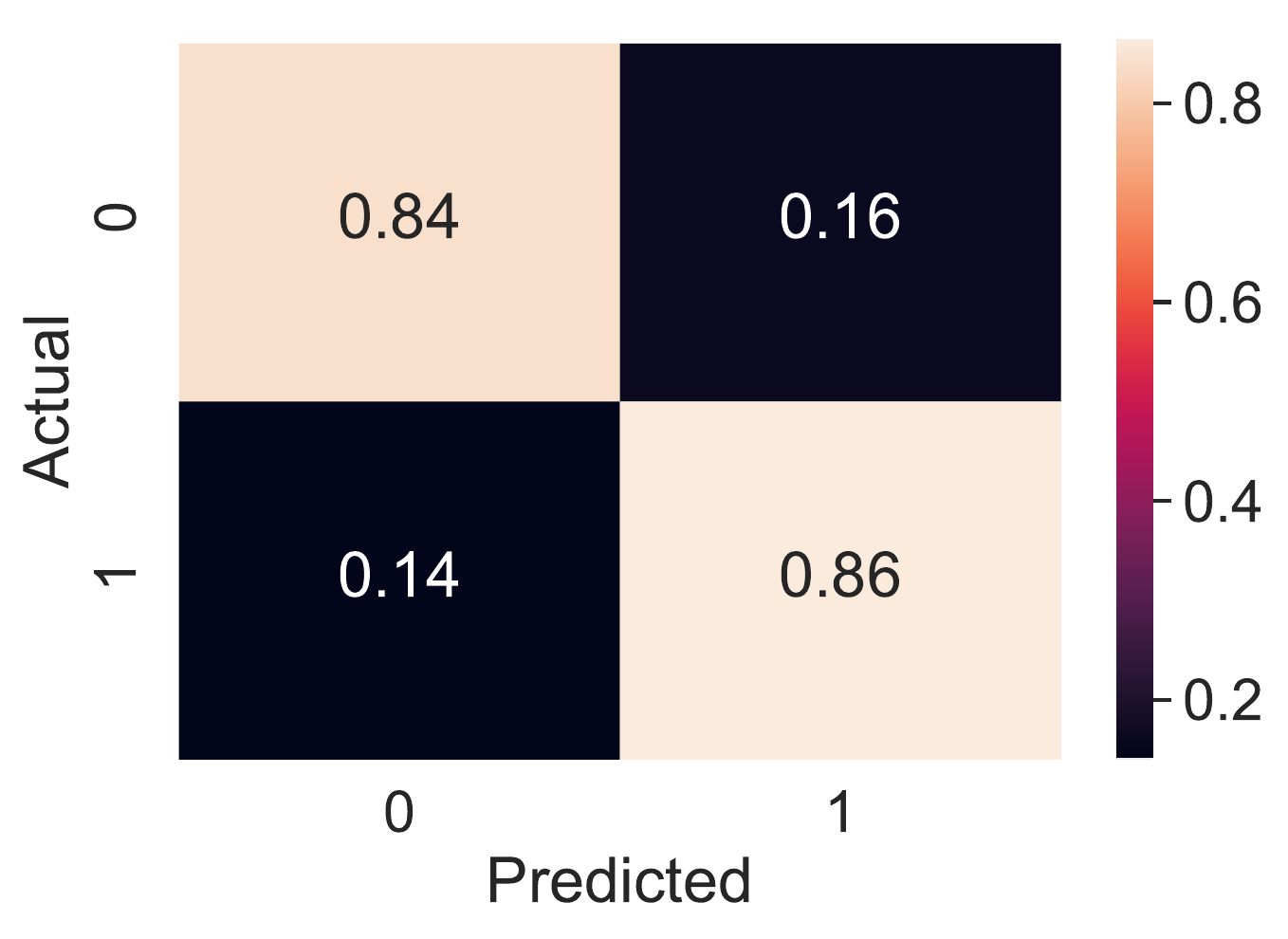}
\includegraphics[height=3.5cm]{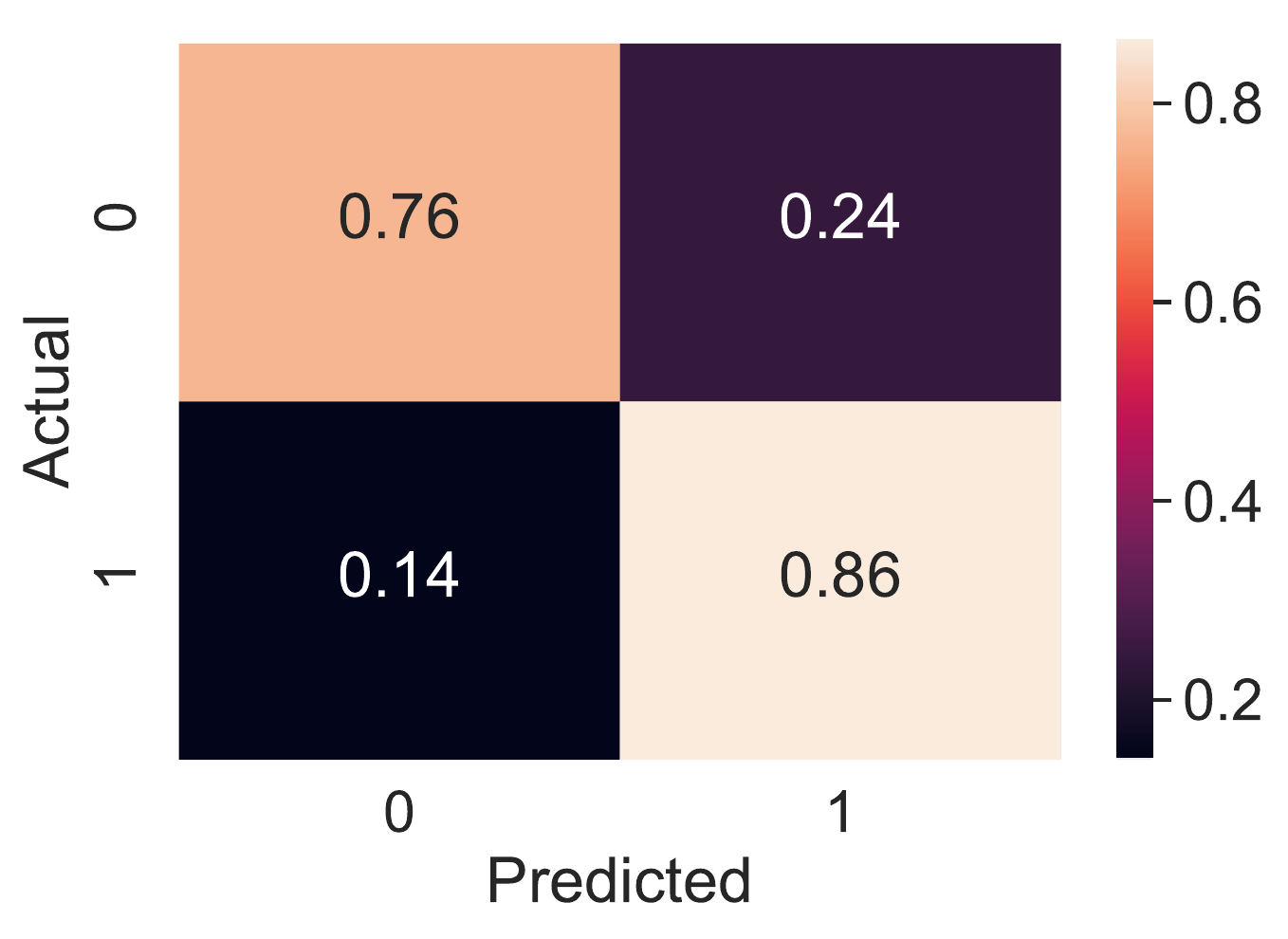}
\caption{\textcolor{black}{Dataset 1: Confusion matrices for the best results of \emph{AutoThorax}-Net (left) and CheXNet (right).}}
\label{fig:confmat1}
\end{figure*}

Experimental results on Dataset 1 are summarized in Table \ref{tab:dataset1_results} \textcolor{black}{for \emph{AutoThorax}-Net, ChexNet and all three feature configurations from Figure \ref{fig:configuration}. We calculated area under the curve (AUC), sensitivity and specificity for all 10 folds. Standard deviations were quite low ($<1\%$), hence not reported. Figure \ref{fig:confmat1} shows the confusion matrices for both \emph{AutoThorax}-Net and ChexNet.}

\begin{table}[ht]
\centering
\textcolor{black}{
\caption {\textcolor{black}{A summary of classification performance using image search as a classifier on Dataset 1. The numbers (in percentage) are the result of avergaing 10 folds with very low standard deviation ($<1\%$).}} 
\label{tab:dataset1_results}
\begin{tabular}{lccc}
\hline
Method &  Sensitivity &  Specificity &  AUC \\
\hline
CheXNet\cite{rajpurkar2017chexnet} classifier &           86 &           76 &       88 \\
\hline
Search via \emph{AutoThorax}-Net features ($k=1001$)  &           86 &           84 &       92 \\
Search via \emph{AutoThorax}-Net features ($k=501$) &           86 &           83 &       92 \\
Search via \emph{AutoThorax}-Net features ($k=251$)  &           84 &           85 &       92 \\
Search via \emph{AutoThorax}-Net features ($k=101$) &           85 &           84 &       92 \\
Search via \emph{AutoThorax}-Net features ($k=51$) &           85 &           84 &       92 \\
Search via \emph{AutoThorax}-Net features ($k=11$) &           81 &           86 &       90 \\
\hline
 Search via Configuration 3 (3072 features, $k=1001$) & 85 & 74 & 88 \\
 Search via Configuration 3 (3072 features, $k=501$) &           84 &           76 &       88 \\
 Search via Configuration 3 (3072 features, $k=251$) &           81 &           79 &       89 \\
 Search via Configuration 3 (3072 features, $k=101$) &           84 &           78 &       89 \\
 Search via Configuration 3 (3072 features, $k=51$) &           86 &           77 &       89 \\
 Search via Configuration 3 (3072 features, $k=11$) &           83 &           80 &       88 \\
\hline     
 Search via Configuration 2 (2048 features, $k=1001$) &           86 &           70 &       87 \\
 Search via Configuration 2 (2048 features, $k=501$) &           85 &           73 &       88 \\
 Search via Configuration 2 (2048 features, $k=251$) &           84 &           75 &       88 \\
 Search via Configuration 2 (2048 features, $k=101$) &           84 &           77 &       88 \\
 Search via Configuration 2 (2048 features, $k=51$) &           79 &           81 &       88 \\
 Search via Configuration 2 (2048 features, $k=11$) &           80 &           81 &       87 \\
\hline      
 Search via Configuration 1 (1024 features, $k=1001$)  &           80 &           78 &       88 \\
 Search via Configuration 1 (1024 features, $k=501$)  &           81 &           78 &       88 \\
 Search via Configuration 1 (1024 features, $k=251$)  &           80 &           80 &       89 \\
 Search via Configuration 1 (1024 features, $k=101$)  &           81 &           80 &       89 \\
 Search via Configuration 1 (1024 features, $k=51$)  &           83 &           80 &       89 \\
 Search via Configuration 1 (1024 features, $k=11$)  &           86 &           76 &       88 \\
\bottomrule
\end{tabular}
}
\end{table}

\textcolor{black}{The average sensitivity and specificity obtained by Configuration 3 for $K=1001$ are higher than those obtained by Configuration 1 although they have almost the same AUC. Configuration 1 shows higher sensitivity for $K=11$ (86\% versus 83\%) but its specificity is lower than Configuration 3 (76\% versus 80\%).
Configuration 2 delivers the same AUC in range 88\% but in individual comparison is always worse than other configurations with lower sensitivity and specificity.}

\textcolor{black}{\emph{AutoThorax}-Net has clearly the highest AUC (92\%). ChexNet delivers an AUC of 88\% similar to the three search configurations. The highest sensitivity is 86\% achieved by all tested methods. However, \emph{AutoThorax}-Net also provides a specificity of 84\% whereas the specificity of all other methods, including ChexNet, are in the 70\% range.}

\subsection*{Second Experiment Series: Automated Solution}
In these experiments, we investigated the possibility of constructing \textcolor{black}{a ``fully automated'' solution by searching the entire archive, i.e., Dataset 2.} We  summarize the experimental workflow, and report the results with some analysis.


\subsubsection*{Experimental Workflow}
\begin{figure*}[htb]
\centering
\includegraphics[height=5.5cm]{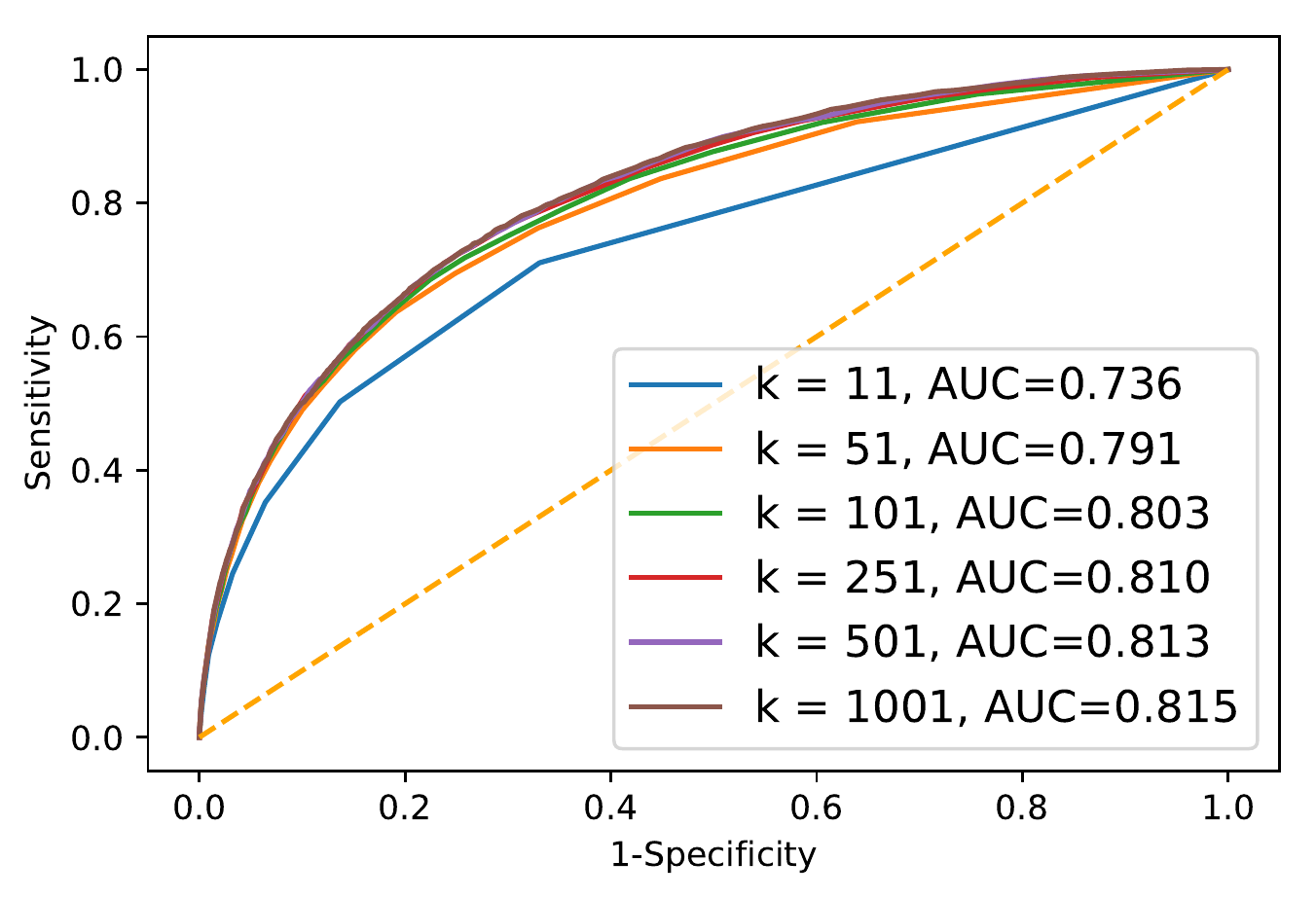}
\includegraphics[height=5.5cm]{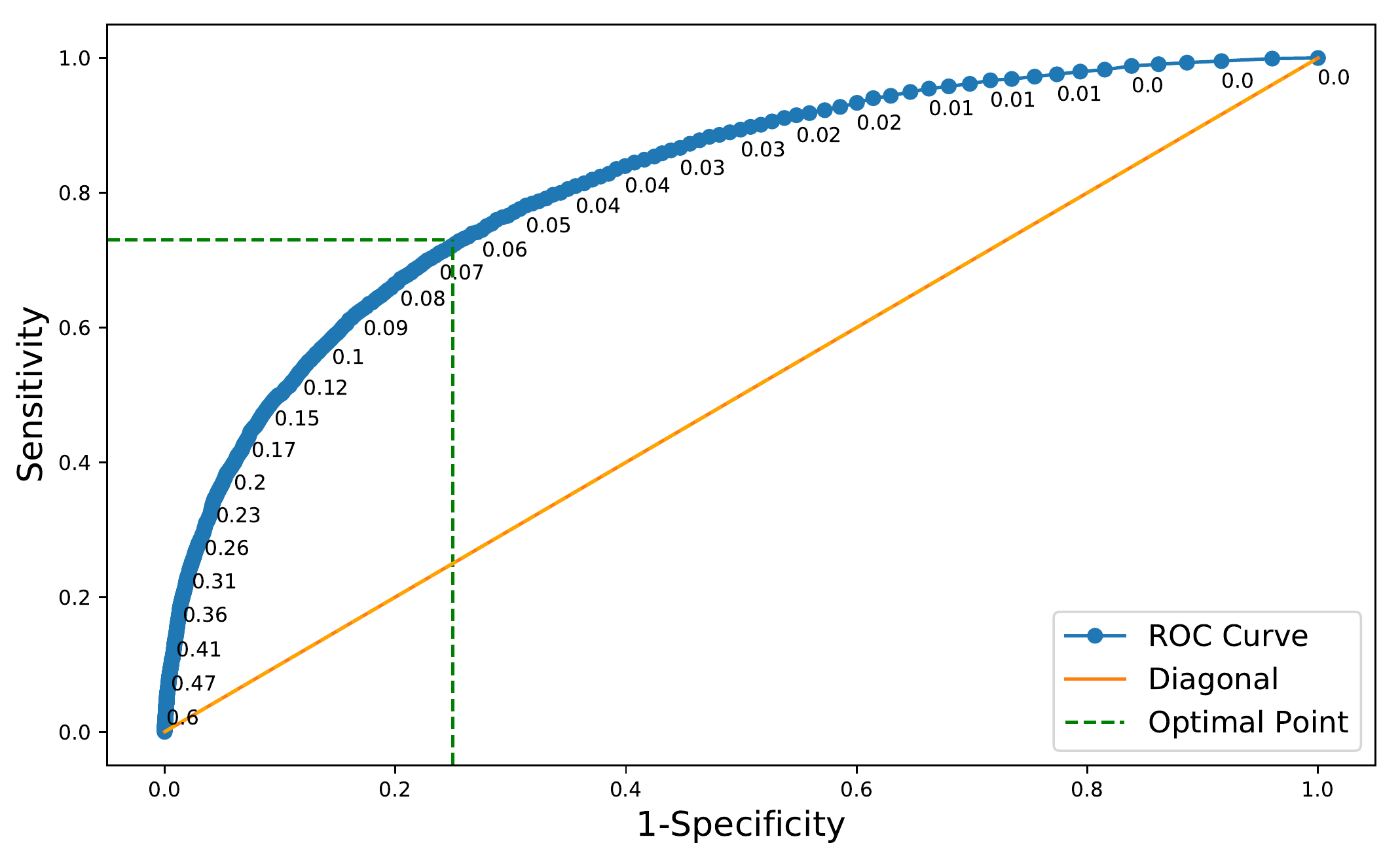}
\caption{\textcolor{black}{Analysis for Dataset 2. Left: Sample ROC curves for the proposed \emph{AutoThorax}-Net for different $k$ values in one fold and their area under the curve (AUC), Right:  corresponding ROC thresholds for $k=1001$ to select the sensitivity-specificity trade-off using Youden's index.}}
\label{fig:roc2}
\end{figure*}

\textcolor{black}{We constructed the receiver operating characteristics (ROC) curve for Dataset 2 to find the trade-off between sensitivity and specificity   (Figure \ref{fig:roc2}). We used Youden's index to find the trade-off position on the ROC curve providing the threshold for match selection.} A standard 10-fold cross-validation was adopted for testing \textcolor{black}{ that showed a very low standard deviation ($<1\%$) for all experiments apparently due to the large size of the datasets}. All chest X-ray images were divided into 10 folds. In each fold, one group of chest X-ray images was used as validation set, while the remaining chest X-ray images were used as archived set.
The above process was repeated 10 times, such that in each fold a different group of chest X-ray images was used as the validation set. In each fold, an encoder was trained using the archived set of that fold. The encoder was then used for compressing deep features for each chest X-ray image in the validation set.

\begin{figure*}[htb]
\centering
\includegraphics[height=3.5cm]{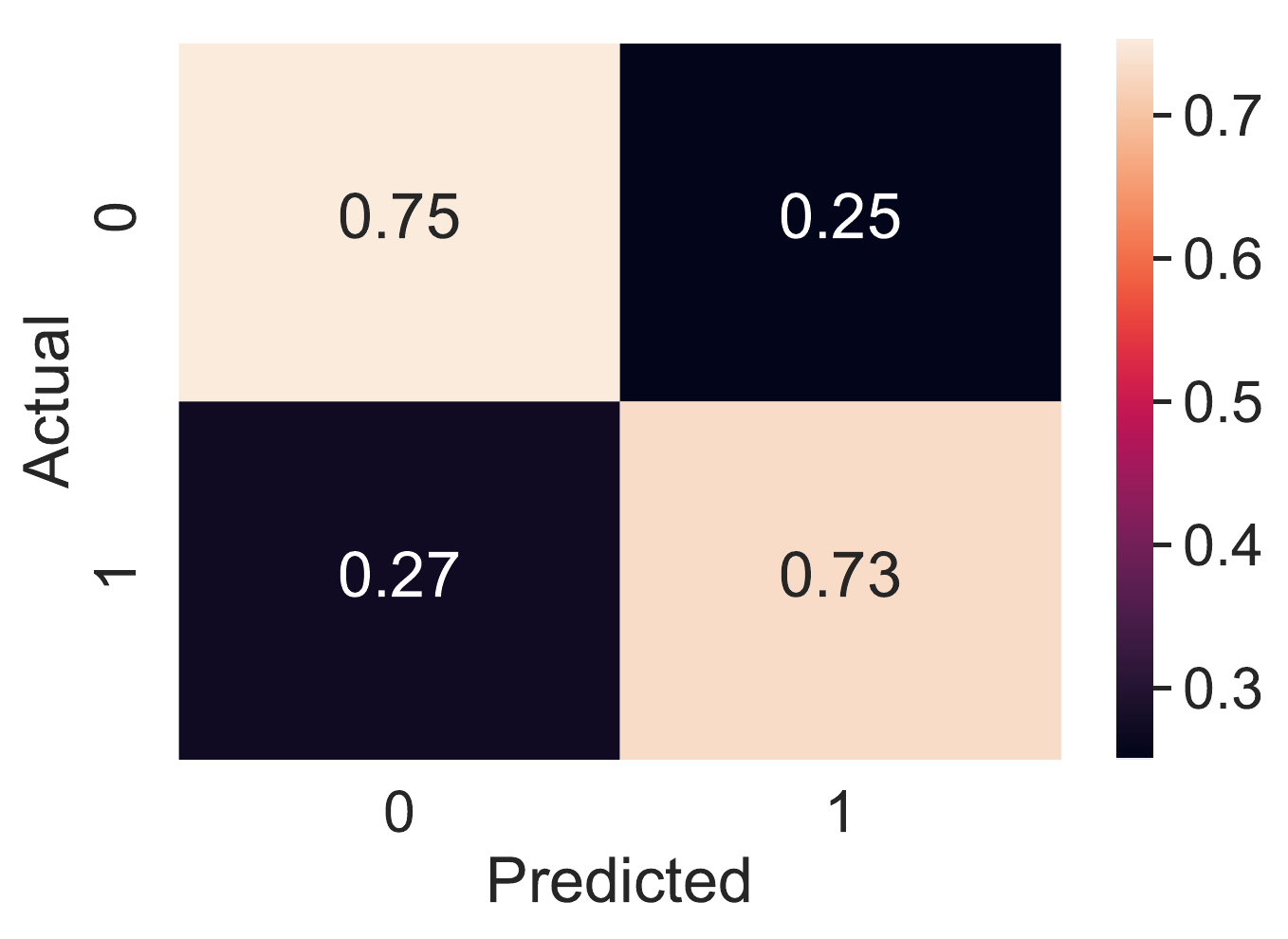}
\includegraphics[height=3.5cm]{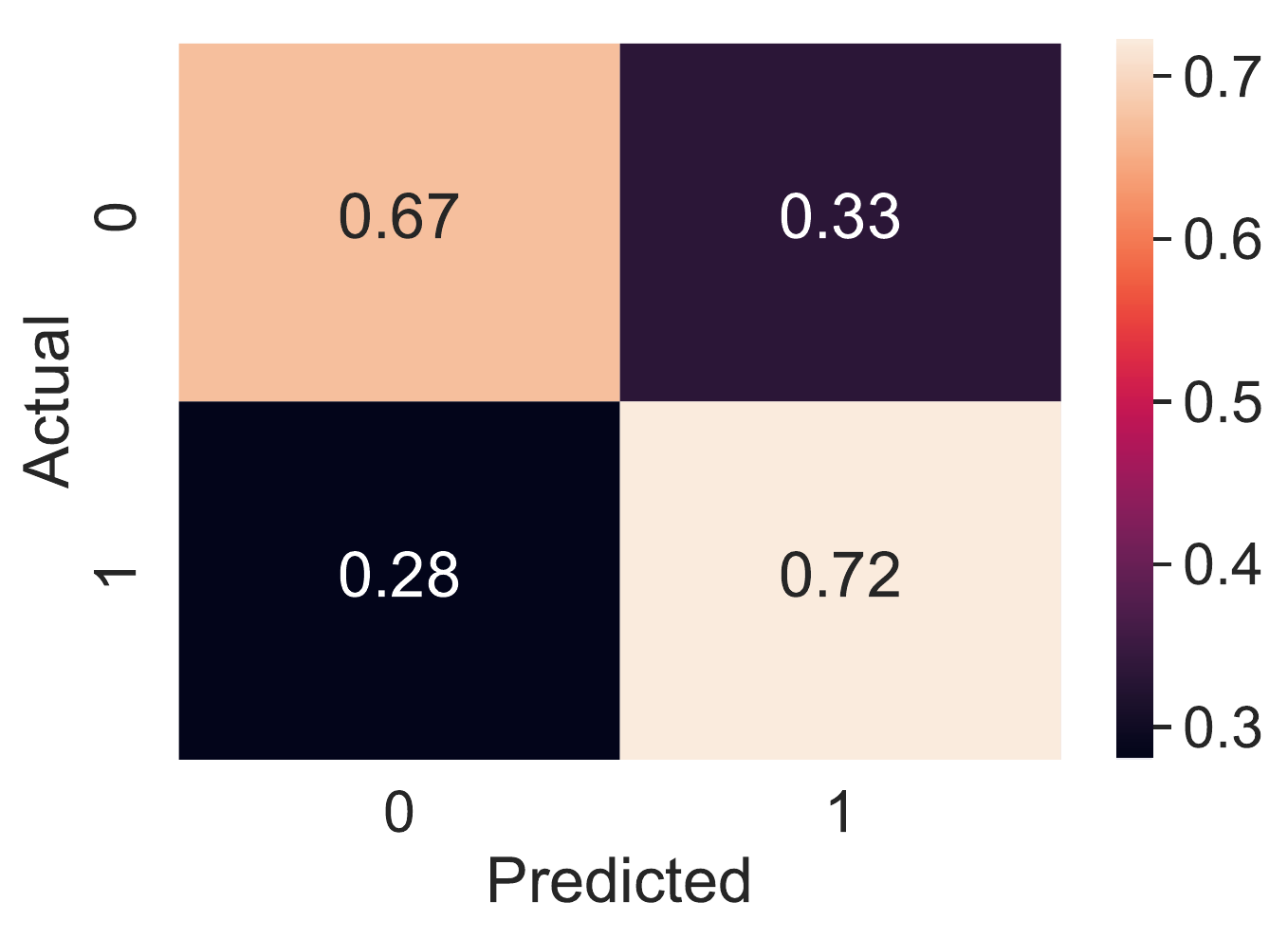}
\caption{\textcolor{black}{Dataset 2: Confusion matrices for the best results of \emph{AutoThorax}-Net (left) and CheXNet (right).}}
\label{fig:confmat2}
\end{figure*}

The parameters of the encoder construction process were set as before described for Dataset 1.  

For image search, given a chest X-ray image (from the validation set), the compressed deep feature was used for searching in the archived set. The consensus vote among the top $K$ retrieved chest X-ray images to classify the query image from the validation set. \textcolor{black}{ Experiments were conducted with $K\in\{11, 51, 101, 251, 501, 1001\}$ to observe the effect of more retrievals on consensus voting.} For comparison, CheXNet \cite{rajpurkar2017chexnet} was adopted as a baseline to be applied to the validation set in each fold.

\subsubsection*{Results}
Experimental results on Dataset 2 are summarized in Table \ref{tab:dataset2_results}. \textcolor{black}{Figure \ref{fig:confmat2} shows the confusion matrices for \emph{AutoThorax}-Net and ChexNet.}

\textcolor{black}{The highest AUC of 82\% is achieved by \emph{AutoThorax}-Net for $k=$ 251, 501 and 1001. The highest sensitivity of 74\% is achieved by Configuration 2 (for $K=251$) and Configuation 3 (for $101$). However, they both deliver low specificity values of 61\% and 65\%, respectively. The second highest sensitivity of 73\% is achieved by Configuration 1, Configuration 2 and \emph{AutoThorax}-Net. Their specificity is 61\%, 63\% and 75\%, respectively. \emph{AutoThorax}-Net can clearly provide a higher and more reliable trade-off between sensitivity and specificity in a fully automated setting when applied on a large archive of X-ray images. }

\begin{table}[ht]
\centering
\textcolor{black}{
\caption {\textcolor{black}{A summary of classification performance using image search as a classifier on Dataset 2. The numbers (in percentage) are the result of avergaing 10 folds with very low standard deviation ($<1\%$).}} 
\label{tab:dataset2_results}
\begin{tabular}{lccc}
\hline
Method &  Sensitivity &  Specificity &  AUC \\
\hline
CheXNet\cite{rajpurkar2017chexnet} Classifier &           72 &           67 &       77 \\
\hline
Search via \emph{AutoThorax}-Net features ($k=1001$) &           73 &           75 &       82 \\
Search via \emph{AutoThorax}-Net features ($k=501$) &           73 &           75 &       82 \\
Search via \emph{AutoThorax}-Net features ($k=251$) &           72 &           75 &       82 \\
Search via \emph{AutoThorax}-Net features ($k=101$) &           69 &           78 &       81 \\
Search via \emph{AutoThorax}-Net features ($k=51$) &           70 &           75 &       80 \\
Search via \emph{AutoThorax}-Net features ($k=11$) &           72 &           67 &       74 \\
\hline
 Search via Configuration 3 (3072 features, $k=1001$) &           72 &           63 &       75 \\
 Search via Configuration 3 (3072 features, $k=501$)  &           70 &           67 &       76 \\
 Search via Configuration 3 (3072 features, $k=251$)  &           71 &           67 &       76 \\
 Search via Configuration 3 (3072 features, $k=101$)  &           74 &           65 &       77 \\
 Search via Configuration 3 (3072 features, $k=51$)  &           65 &           74 &       76 \\
 Search via Configuration 3 (3072 features, $k=11$) &           72 &           65 &       72 \\
\hline
 Search via Configuration 2 (2048 features, $k=1001$) &           67 &           66 &       74 \\
 Search via Configuration 2 (2048 features, $k=501$) &           64 &           70 &       75 \\
 Search via Configuration 2 (2048 features, $k=251$) &           74 &           61 &       75 \\
 Search via Configuration 2 (2048 features, $k=101$) &           70 &           66 &       75 \\
 Search via Configuration 2 (2048 features, $k=51$) &           73 &           63 &       75 \\
 Search via Configuration 2 (2048 features, $k=11$)&           68 &           66 &       70 \\
\hline      
 Search via Configuration 2 (1024 features, $k=1001$) &           73 &           61 &       74 \\
 Search via Configuration 2 (1024 features, $k=501$) &           67 &           68 &       75 \\
 Search via Configuration 2 (1024 features, $k=251$) &           67 &           69 &       75 \\
 Search via Configuration 2 (1024 features, $k=101$) &           70 &           65 &       75 \\
 Search via Configuration 2 (1024 features, $k=51$) &           67 &           68 &       74 \\
 Search via Configuration 2 (1024 features, $k=11$) &           71 &           60 &       69 \\
\bottomrule
\end{tabular}
}
\end{table}


\subsection*{\textcolor{black}{Comparing Autoencoder against PCA}}
\textcolor{black}{As one of the main contributions of the \emph{AutoThorax}-Net is encoding the concatenated feature vector (i.e., reducing the dimensionality), the question arises whether the same level of performance can be achieved by traditional algorithms such as the principal component analysis (PCA). We did run the 10-fold cross validation on both dataset configurations for $k=11$ and $k=51$. We observed in all settings that the performance of autoencoder was better than PCA. For instance, for the second experiment, PCA achieved 72\% and 76\% AUC for $k=11$ and $k=51$, respectively, while autoencoder achieved 74\% and 80\% AUC. As the performance of the dimensionality reduction is independent of $k$, one expects that a more capable compression should already manifest itself for any $k$. However, as we are using the compressed/encoded features for image search, good performance is expected to be particularly visible for a small number of matched cases. }

\section*{Discussions}
In our investigations, we experimented with image search as a classifier to detect pneumothorax based on autoencoded concatenated features applied on more than half a million chest X-ray images obtained through the merging three large public datasets. 

In our experiments, we verified that the use of image search as a classifier with \emph{AutoThorax}-Net as a feature extractor can improve classification performance.  This was demonstrated by analysing the ROC curves to find the trade-off for each individual approach. We further confirmed that compressing concatenated deep features via autoencoders further improves the results of image search. This indicates that image search as a classifier is a viable and more conveniently explainable solution for the practice of diagnostic radiology when reports and history of evidently diagnosed cases of similar cases are readily available.

\section*{Acknowledgements}
This research was a ``Path-Finder Project'' financially supported by the Vector Institute, Toronto, Canada.



\bibliography{paper}

\end{document}